\documentclass[global]{svjour}
\usepackage{amssymb}
\usepackage{amsfonts}
\usepackage{amsmath}

\input{tcilatex}
\begin{document}

\journalname{AHEP 739153}
\title{Resonance Spectra of Caged Stringy Black Hole and Its Spectroscopy}
\author{I. Sakalli\inst{1} \and G.Tokgoz\inst{2}}
\institute{Department of Physics, Eastern Mediterranean University, Gazimagosa, North
Cyprus, Mersin 10, Turkey, \ 
\email{izzet.sakalli@emu.edu.tr}%
.\\
\and Department of Physics, Eastern Mediterranean University, Gazimagosa,
North Cyprus, Mersin 10, Turkey, \ 
\email{gulnihal.tokgoz@emu.edu.tr}%
.}
\dedication{}
\offprints{}
\mail{}
\maketitle

\begin{abstract}
The Maggiore's method (MM), which evaluates the transition frequency that
appears in the adiabatic invariant from the highly damped quasinormal mode
(QNM) frequencies, is used to investigate the entropy/area spectra of the
Garfinkle--Horowitz--Strominger black hole (GHSBH). Instead of the ordinary
QNMs, we compute the boxed QNMs (BQNMs) that are the characteristic
resonance spectra of the confined scalar fields in the GHSBH geometry. For
this purpose, we assume that the GHSBH has a confining cavity (mirror)
placed in the vicinity of the event horizon. We then show how the complex
resonant frequencies of the caged GHSBH are computed using the Bessel
differential equation that arises when the scalar perturbations around the
event horizon are considered. Although the entropy/area is characterized by
the GHSBH parameters, their quantization is shown to be independent of those
parameters. However, both spectra are equally spaced.
\end{abstract}


\section{Introduction}

Currently, one of the greatest projects in theoretical physics is to unify
general relativity (GR) with quantum mechanics (QM). Such a new unified
theory is known as the quantum gravity theory (QGT) \cite{Robson}. Recent
developments in physics show that our universe has a more complex structure
than that predicted by the standard model \cite{Oerter}. The QGT is
considered to be an important tool that can tackle this problem. However,
current QGT still requires further extensive development to reach
completion. The development of the QGT began in the seventies when Hawking 
\cite{Hawking1,Hawking2} and Bekenstein \cite{Bek1,Bek2,Bek3,Bek4,Bek5}
considered the black hole (BH) as a quantum mechanical system rather than a
classical one. In particular, Bekenstein showed that the area of the BH
should have a discrete and equally spaced spectrum:

\begin{equation}
\mathcal{A}_{n}=\epsilon n\hbar =8\pi \xi n\hbar \text{ \ \ \ \ \ \ }%
(n=0,1,2.......),  \label{1}
\end{equation}

where $\epsilon $ is the undetermined dimensionless constant and $\xi $\ is
of the order of unity. The above expression also shows that the minimum
increase in the horizon area is $\Delta \mathcal{A}_{\min }=\epsilon \hbar $%
. Bekenstein \cite{Bek3,Bek4} also conjectured that for the Schwarzschild BH
(and also for the Kerr-Newman BH) the value of $\epsilon $ is $8\pi $ (or $%
\xi =1$). Following the seminal work of Bekenstein, various methods have
been suggested to compute the area spectrum of the BHs. Some methods used
for obtaining the spectrum can admit the value of $\epsilon $ different than
that obtained by Bekenstein; this has led to the discussion of this subject
in the literature (for a review of this topic, see \cite{Gua} and references
therein). Among those methods, the MM's results \cite{Maggiore} show a
perfect agreement with Bekenstein's result by modifying the Kunstatter's 
\cite{Kunstatter} formula as

\begin{equation}
I_{adb}=\int \frac{dM}{\Delta \omega },  \label{2}
\end{equation}

where $I_{adb}$ denotes the adiabatic invariant quantity and $\Delta \omega
=\omega _{n-1}-\omega _{n}$ represents the transition frequency between the
subsequent levels of an uncharged and static BH with the total energy (or
mass) $M$. However, the researchers \cite{Vagenas,Medved,Sakalli14} working
on this issue later realized that the above definition is not suitable for
the charged rotating (hairy) BHs that the generalized form of the definition
should be given by

\begin{equation}
I_{adb}=\int \frac{TdS}{\Delta \omega },  \label{3}
\end{equation}

where $T$ and $S$ denote the temperature and the entropy of the BH,
respectively. Thus, using the first law of BH thermodynamics, Eq. (3) can be
modified for the considered BH. According to the Bohr--Sommerfeld
quantization rule, $I_{adb}$\ behaves as a quantized quantity $%
(I_{adb}\simeq n\hbar )$ while the quantum number $n$ tends to infinity. To
determine $\Delta \omega $, Maggiore considered the BH as a damped harmonic
oscillator that has a proper physical frequency in the form of $\omega
=\left( \omega _{R}^{2}+\omega _{I}^{2}\right) ^{\frac{1}{2}}$, where $%
\omega _{R}$\ and $\omega _{I}$\ are the real and imaginary parts of the
frequency, respectively. For the highly excited modes ($n\rightarrow \infty $%
), $\omega _{I}\gg \omega _{R}$ and therefore $\Delta \omega \simeq \Delta
\omega _{I}$. Hod \cite{Hod1,Hod2} was the first to argue that the QNMs can
be used in the identification of the quantum transitions for the $I_{adb}$.
Subsequently, there have been other published papers that use the MM to
achieve similar results (see for instance \cite%
{Samp1,Samp2,Samp3,Samp4,Samp5,Samp6,Samp7,Samp8}).

In this paper, we focus on the investigation of the GHSBH \cite{GHS}
spectra. The GHSBH is a member of a family of solutions for the low-energy
limit of the string theory. This spacetime is obtained when the field
content of the Einstein--Maxwell theory is enlarged to cover a dilaton
field, which couples to the metric and the gauge field non-trivially. This
causes the charged stringy BHs to differ significantly from the
Reissner--Nordstr\"{o}m (RN) BH. To employ the MM, the QNMs (a set of
complex frequencies arising from the perturbed BH) of the GHSBH should be
computed. To achieve this, we first consider the KGE for a massless scalar
field in the background of the GHSBH. After separating the angular and the
radial equations, we obtain a Schr\"{o}dinger-like wave equation, which is
the so-called Zerilli equation \cite{Chandra}.

In fact, the spectroscopy of the GHSBH was previously studied by Wei et al. 
\cite{Wei}. They used the QNMs of Chen and Jing \cite{Chen} who studied the
monodromy method \cite{Motl} and obtained an equal spacing of GHSBH spectra
at the high frequency modes. There are several methods to calculate the
QNMs, such as the WKB method, the phase integral method, continued fractions
and direct integrations of the wave equation in the frequency domain \cite%
{Berti}. Our main goal in this study is to consider a recent analytical
method, which is invented by Hod \cite{Hodnew} for obtaining the GHSBH's
resonance spectra or the BQNMs \cite%
{Cardoso,RanLi1,Herdeiro1,Herdeiro2,RanLi2,RanLi3}, and is different from
the monodromy method. Thus, we seek to support the study of Wei et al. \cite%
{Wei} because we believe that the studies that obtain the same conclusion
using different methods are more reliable. For this purpose, we consider the
GHSBH as a caged BH, which describes a BH\ confined within a finite-volume
cavity. The Hod's idea is indeed based on the recent study \cite{Okawa14},
which provides compelling evidence that spherically-symmetric confined
scalar fields in a cavity of Einstein-Klein-Gordon system generically
collapse to form caged BHs. To this end, we consider a mirror (confining
cavity) surrounding the GHSBH that is placed at a constant radial coordinate
with a radius $r_{m}$. The scalar field $\Phi $ is imposed to terminate at
the mirror's location, which requires to use the Dirichlet boundary
condition (DBC) ($\left. \Phi (r)\right\vert _{r=r_{m}}=0$) and Neumann
boundary condition (NBC) ($\left. \frac{d\Phi (r)}{dr}\right\vert
_{r=r_{m}}=0$). In the framework of this scenario, we focus our analysis of
the radial wave equation on the near-horizon (NH) region \cite{Hodnew}. We
then derive the BQNMs of the GHSBH based on the fact that for the QNMs to
exist, the outgoing waves must be terminated at the event horizon. The NH
form of the Zerilli equation is reduced to a Bessel differential equation 
\cite{Abram}. After choosing the expedient solution, we impose the Dirichlet
and Neumann boundary conditions. Then, we consider some of the transformed
features of the Bessel functions for finding the resonance conditions. Next,
we use an iteration scheme to define the BQNMs of the GHSBH. Once the BQNMs
are obtained, we use the transition frequency $\Delta \omega _{I}$ in Eq.
(3) and obtain the GHSBH area/entropy spectra.

The remainder of this paper is arranged as follows. Sec. 2 introduces the
GHSBH metric, and analyzes the KGE for a massless scalar field in this
geometry. Using the separation of variables technique, we then reduce the
physical problem to the Zerilli equation. In Sec. 3, we show that the
Zerilli equation reduces to a Bessel differential equation in the vicinity
of the event horizon. The Dirichlet and Neumann boundary conditions at the
surface $r=r_{m}$ of the confining cavity single out two discrete sets of
complex BQNMs of the caged GHSBH. Finally, we apply the MM to obtain the
quantum spectra of the entropy/area of the GHSBH. Conclusions are presented
in Sec. 4. (Throughout the paper, we set $c=G=k_{B}=1$).

\section{GHSBH and the Separation of the Massless KGE on It}

In this section, we represent the geometry and some of the thermodynamical
properties of the GHSBH \cite{GHS}. We also derive the Zerilli equation and
its corresponding effective potential for a massless scalar field
propagating in the GHSBH background.

In the low-energy limit of the string field theory, the four-dimensional
Einstein-Maxwell-dilaton low-energy action (in Einstein frame) describing
the dilaton field $\phi $ coupled to a $U(1)$ gauge field is given by

\begin{equation}
S=\int d^{4}x\sqrt{-g}(-R+2(\nabla \phi )^{2}+e^{-2\phi }F^{2}),  \label{4}
\end{equation}

with $F^{2}=F_{\mu \upsilon }F^{\mu \upsilon }$ in which $F_{\mu \upsilon }$
is the Maxwell field associated with the $U(1)$ subgroup of $E_{8}\times
E_{8}$ or Spin(32)/$Z_{2}$ \cite{GHS}. After applying the variational
principle to the above action, we obtain the following field equations

\begin{equation}
\nabla _{\mu }(e^{-2\phi }F^{\mu \nu })=0,  \label{5}
\end{equation}%
\begin{equation}
\nabla ^{2}\phi +\frac{1}{2}e^{-2\phi }F^{2}=0,  \label{6}
\end{equation}

\begin{equation}
R_{\mu \nu }=2\nabla _{\mu }\phi \nabla _{\nu }\phi -g_{\mu \nu }\left(
\nabla \phi \right) ^{2}+2e^{-2\phi }F_{\mu \rho }F_{\nu }^{\rho }-\frac{1}{2%
}g_{\mu \nu }e^{-2\phi }F^{2}.  \label{7}
\end{equation}

Their solutions are expressed by the following static and spherically
symmetric metric. 
\begin{equation}
ds^{2}=-f(r)dt^{2}+\frac{dr^{2}}{f(r)}+A(r)d\Omega ^{2},  \label{8}
\end{equation}

where $d\Omega ^{2}$ is the standard metric on the $2$-sphere. The metric
functions are given by

\begin{equation}
f(r)=1-\frac{r_{+}}{r},  \label{9}
\end{equation}

\begin{equation}
A(r)=r(r-2a).  \label{10}
\end{equation}

The physical parameter $a$ is defined by

\begin{equation}
a=\frac{Q^{2}e^{-2\phi _{0}}}{2M},  \label{11}
\end{equation}

where $Q$, $M$ and $\phi _{0}$ describe the magnetic charge, mass and the
asymptotic constant value of the dilaton, respectively. Besides, $r_{+}=2M$
represents the event horizon of the GHSBH. In this spacetime, the dilaton
field is governed by

\begin{equation}
e^{-2\phi }=e^{-2\phi _{0}}\left( 1-\frac{2a}{r}\right) ,  \label{12}
\end{equation}

and the Maxwell field reads

\begin{equation}
F=Q\sin \theta d\theta \wedge d\varphi .  \label{13}
\end{equation}

For the electric charge case, one can simply apply the following duality
transformations. 
\begin{equation}
\tilde{F}_{\mu \nu }=\frac{1}{2}e^{-2\phi }\epsilon _{\mu \nu }^{\lambda
\rho }F_{\lambda \rho },\text{ \ \ \ \ \ \ \ \ \ \ }\phi \rightarrow -\phi .
\label{14}
\end{equation}

Since $R^{2}$ part of the GHSBH metric (8) is identical to the Schwarzschild
BH, the surface gravity \cite{Wald} naturally coincides with the
Schwarzschild's one

\begin{equation}
\kappa =\lim_{r\rightarrow r_{+}}\sqrt{-\frac{1}{2}\nabla ^{\mu }\chi ^{\nu
}\nabla _{\mu }\chi _{\nu }}=\left. \frac{f^{\prime }(r)}{2}\right\vert
_{r=r_{+}}=\frac{1}{4M},  \label{15}
\end{equation}

where the timelike Killing vector is $\chi ^{\nu }=[1,0,0,0]$. Therefore,
the Hawking temperature $T_{H}$ of the GHSBH reads

\begin{equation}
T_{H}=\frac{\hbar \kappa }{2\pi }=\frac{\hbar }{8\pi M}.  \label{16}
\end{equation}

Thus, the Hawking temperature of the GHSBH is independent of the amount of
the charge. But, the similarity between the GHSBH and the Schwarzschild BH
is apparent since the radial coordinate does not belong to the areal radius.
So, the entropy of the GHSBH is different than the Schwarzschild BH's
entropy:

\begin{equation}
S^{BH}=\frac{\mathcal{A}}{4\hbar }=\frac{\pi r_{+}(r_{+}-2a)}{\hbar }.
\label{17}
\end{equation}

In fact, at extremal charge $Q=\sqrt{2}Me^{\phi _{0}}$ i.e., $a=M$, the BH
has a vanishing area and hence its entropy is zero. The extremal GHSBH is
not a BH in the ordinary sense since its area has become degenerate and
singular: it is indeed a naked singularity. Unlike to the singularity of RN,
which is timelike, this singularity is null and whence outward-directed
radial null geodesics cannot hit it. For a detailed study of the null
geodesics of the GHSBH, one may refer to \cite{NullGeo}. On the other hand,
one can easily prove that the first law of thermodynamics:

\begin{equation}
T_{H}dS^{BH}=dM-V_{H}dQ,  \label{18}
\end{equation}

holds for the GHSBH. In Eq. (18), the electric potential on the horizon is $%
V_{H}=\frac{a}{Q}$. To obtain the GHSBH spectra via the MM, we shall first
consider the massless scalar field $\Psi $ satisfying the KGE:

\begin{equation}
\frac{1}{\sqrt{-g}}\partial _{\nu }(\sqrt{-g}g^{\mu \nu }\partial _{\mu
}\Psi )=0.  \label{19}
\end{equation}

The chosen ansatz for the scalar field $\Psi $ has the following form

\begin{equation}
\Psi =A(r)^{-1/2}\digamma (r)e^{-i\omega t}Y_{l}^{m}(\theta ,\varphi ),\text{
\ \ }Re(\omega )>0,  \label{20}
\end{equation}

in which $\omega \ $and $Y_{l}^{m}(\theta ,\varphi )$ represent the
frequency of the propagating scalar wave and the spheroidal harmonics with
the eigenvalue $L=-l(l+1),$ respectively. After some algebra, the radial
equation can be reduced to the following form

\begin{equation}
\left[ -\frac{d^{2}}{dr^{\ast 2}}+V(r)\right] \digamma (r)=\omega
^{2}\digamma (r),  \label{21}
\end{equation}

which is nothing but the Zerilli equation \cite{Chandra}. Employing the
tortoise coordinate $r^{\ast }$ defined as

\begin{equation}
r^{\ast }=\dint \frac{dr}{f(r)},  \label{22}
\end{equation}

we get

\begin{equation}
r^{\ast }=r+r_{+}\ln \left( \frac{r}{r_{+}}-1\right) .  \label{23}
\end{equation}

Inversely, one obtains

\begin{equation}
r=r_{+}\left[ 1+W\left( u\right) \right] ,  \label{24}
\end{equation}

where $u=e^{\left( \frac{r^{\ast }}{r_{+}}-1\right) }$ and $W\left( u\right) 
$ represents the Lambert-W or the omega function \cite{Lambert}. It can be
checked that

\begin{equation}
\lim_{r\rightarrow r_{+}}r^{\ast }=-\infty ,\text{ \ and\ \ \ }%
\lim_{r\rightarrow \infty }r^{\ast }=\infty .  \label{25}
\end{equation}

The effective or the so-called Zerilli potential $V(r)$\ is given by

\begin{equation}
V(r)=\frac{f(r)}{r\left( r-2a\right) }\left[ L-\frac{a^{2}}{r\left(
r-2a\right) }f(r)+\frac{2M(r-a)}{r^{2}}\right] .  \label{26}
\end{equation}

\section{BQNM Frequencies and Entropy/Area Spectra}

In this section, we interest in solutions of the Zerilli equation (20)
around the NH. In computing the BQNM frequencies, we first impose the
condition that QNMs should be ingoing plane waves at the event horizon.
Secondly, we borrow ideas from recent study \cite{Hodnew} and impose the DBC
and NBC to have the resonance conditions. In computing the BQNMs, we use an
iteration scheme to resolve the resonance conditions.

The metric function $f(r)$ can be rewritten as follows

\begin{equation}
f(r)\rightarrow f(x)=\frac{x}{x+1},  \label{27}
\end{equation}

where

\begin{equation}
x=\frac{r-r_{+}}{r_{+}}.  \label{28}
\end{equation}

Thus, one finds

\begin{equation}
f(x)\cong x+O(x^{2}),  \label{29}
\end{equation}

\begin{equation}
r^{\ast }=\dint \frac{r_{+}dx}{f(x)}\cong r_{+}\ln (x)=\frac{1}{2\kappa }\ln
x,  \label{30}
\end{equation}

in the NH region ($x\rightarrow 0$). From Eq. (30), one reads

\begin{equation}
x=e^{2y},  \label{31}
\end{equation}

where

\begin{equation}
y=\kappa r^{\ast }.  \label{32}
\end{equation}

After substituting Eq. (28) into Eq. (26), the NH form of the Zerilli
potential can be approximated by

\begin{equation}
V_{NH}(x)=\frac{L+\beta }{r_{+}\gamma }x+O(x^{2}),  \label{33}
\end{equation}

where the parameters\ are\ given by

\begin{equation}
\beta =\frac{r_{+}-a}{r_{+}}\text{ \ \ \ ;\ \ \ \ \ }\gamma =r_{+}-2a.
\label{34}
\end{equation}

Substituting Eqs. (31-33) into Eq. (21), one obtains the following NH\ form
of the Zerilli equation

\begin{equation}
\left[ -\frac{d^{2}}{dy^{2}}+\frac{4r_{+}\left( L+\beta \right) }{\gamma }%
e^{2y}\right] \digamma (y)=\widetilde{\omega }^{2}\digamma (y).  \label{35}
\end{equation}

The above differential equation has two linearly independent solutions:

\begin{equation}
\digamma (y)=C_{1}J_{-i\widetilde{\omega }}(2i\sqrt{\Delta }e^{y})+C_{2}Y_{-i%
\widetilde{\omega }}(2i\sqrt{\Delta }e^{y}),  \label{36}
\end{equation}

and correspondingly

\begin{equation}
\digamma (x)=C_{1}J_{-i\widetilde{\omega }}(2i\sqrt{\Delta x})+C_{2}Y_{-i%
\widetilde{\omega }}(2i\sqrt{\Delta x}),  \label{37}
\end{equation}

where $J_{\upsilon }(z)$ and $Y_{\upsilon }(z)$ are called Bessel functions 
\cite{Abram} of the first and second kinds, respectively. $C_{1},$ $C_{2}$
are constants. The parameters of the special functions are given by

\begin{equation}
\widetilde{\omega }=\frac{\omega }{\kappa },  \label{38}
\end{equation}

\begin{equation}
\Delta =\frac{r_{+}\left( L+\beta \right) }{\gamma }.  \label{39}
\end{equation}

The following limiting forms (when $\upsilon $ is fixed and $z\rightarrow 0$%
) of the Bessel functions are needed for our analysis \cite{Abram,Olver}.

\begin{equation}
J_{\upsilon }(z)\sim \frac{\left( \frac{1}{2}z\right) ^{\upsilon }}{\Gamma
(1+\upsilon )},\text{ \ \ \ \ \ \ \ }\left( \upsilon \neq
-1,-2,-3,...\right) ,  \label{40}
\end{equation}

\begin{equation}
Y_{\upsilon }(z)\sim -\frac{1}{\pi }\Gamma (\upsilon )\left( \frac{1}{2}%
z\right) ^{-\upsilon },\text{ \ }(\Re \upsilon >0).  \label{41}
\end{equation}

By using them, we obtain the NH ($e^{y}\ll 1$) behavior of the solution (36)
as

\begin{eqnarray}
\digamma &\sim &C_{1}\frac{\left( i\sqrt{\Delta }\right) ^{-i\widetilde{%
\omega }}}{\Gamma (1-i\widetilde{\omega })}e^{-i\widetilde{\omega }y}-C_{2}%
\frac{1}{\pi }\Gamma (-i\widetilde{\omega })\left( i\sqrt{\Delta }\right) ^{i%
\widetilde{\omega }}e^{i\widetilde{\omega }y},  \notag \\
&=&C_{1}\frac{\left( i\sqrt{\Delta }\right) ^{-i\widetilde{\omega }}}{\Gamma
(1-i\widetilde{\omega })}e^{-i\omega r^{\ast }}-C_{2}\frac{1}{\pi }\Gamma (-i%
\widetilde{\omega })\left( i\sqrt{\Delta }\right) ^{i\widetilde{\omega }%
}e^{i\omega r^{\ast }},  \label{42}
\end{eqnarray}

in which $C_{1}$ and $C_{2}$ correspond to the amplitudes of the NH ingoing
and outgoing waves, respectively. Since the QNMs impose that the outgoing
waves must spontaneously terminate at the horizon, we deduce that $C_{2}=0$.
Thus, the acceptable solution of the radial equation (37) is given by

\begin{equation}
\digamma (x)=C_{1}J_{-i\widetilde{\omega }}(2i\sqrt{\Delta x}).  \label{43}
\end{equation}

Following \cite{Hodnew,Okawa14}, we consider the DBC at the surface $x=x_{m}$%
\ of the confining cage:

\begin{equation}
\left. \digamma (x)\right\vert _{x=x_{m}}=0.  \label{44}
\end{equation}

Thus, we have

\begin{equation}
J_{-i\widetilde{\omega }}(2i\sqrt{\Delta x_{m}})=0.  \label{45}
\end{equation}

Using the following relation \cite{Abram}

\begin{equation}
Y_{\upsilon }(z)=J_{\upsilon }(z)\cot (\upsilon \pi )-J_{-\upsilon }(z)\csc
(\upsilon \pi ),  \label{46}
\end{equation}

we can express the condition (45) as

\begin{equation}
\tan (i\widetilde{\omega }\pi )=\frac{J_{i\widetilde{\omega }}(2i\sqrt{%
\Delta x_{m}})}{Y_{i\widetilde{\omega }}(2i\sqrt{\Delta x_{m}})},  \label{47}
\end{equation}

which is called the resonance condition. According to the definition of the
caged BHs, the boundary of the confining cavity is located at the vicinity
of the event horizon \cite{Hodnew}. Namely, we have

\begin{equation}
z_{m}\equiv \Delta x_{m}\ll 1,\text{ \ \ \ \ }\rightarrow \text{ \ \ \ \ }%
r_{m}\approx r_{+},\text{\ }  \label{48}
\end{equation}

in the NH\ region. Using Eqs. (40) and (41), we can rewrite the resonance
condition (47) as

\begin{eqnarray}
\tan (i\widetilde{\omega }\pi ) &\sim &-\frac{\pi \left( i\sqrt{z_{m}}%
\right) ^{i\widetilde{\omega }}}{\Gamma \left( i\widetilde{\omega }\right)
\Gamma \left( i\widetilde{\omega }+1\right) \left( i\sqrt{z_{m}}\right) ^{-i%
\widetilde{\omega }}},  \notag \\
&=&i\frac{\pi e^{-\pi \widetilde{\omega }}}{\widetilde{\omega }\Gamma
^{2}\left( i\widetilde{\omega }\right) }z_{m}^{i\widetilde{\omega }}.
\label{49}
\end{eqnarray}

The NBC is given by \cite{Hodnew,Okawa14}

\begin{equation}
\left. \frac{d\digamma (x)}{dx}\right\vert _{x=x_{m}}=0.  \label{50}
\end{equation}

Using the derivative features of the Bessel functions given in the monograph 
\cite{Abram}, we obtain

\begin{equation}
J_{-i\widetilde{\omega }-1}(2i\sqrt{z_{m}})-J_{-i\widetilde{\omega }+1}(2i%
\sqrt{z_{m}})=0.  \label{51}
\end{equation}

Using Eq. (46), we derive the following relation

\begin{eqnarray}
Y_{\upsilon +1}(z)-Y_{\upsilon -1}(z) &=&\cot (\upsilon \pi )\left[
J_{\upsilon +1}(z)-J_{\upsilon -1}(z)\right]  \notag \\
&&-\csc (\upsilon \pi )\left[ J_{-\upsilon -1}(z)-J_{-\upsilon +1}(z)\right]
.  \label{52}
\end{eqnarray}

Combining Eqs. (51) and (52), we further get the NBC's resonance condition:

\begin{equation}
\tan (i\widetilde{\omega }\pi )=\frac{J_{i\widetilde{\omega }-1}(2i\sqrt{%
z_{m}})}{Y_{i\widetilde{\omega }+1}(2i\sqrt{z_{m}})}\left[ \frac{-1+\frac{%
J_{i\widetilde{\omega }+1}(2i\sqrt{z_{m}})}{J_{i\widetilde{\omega }-1}(2i%
\sqrt{z_{m}})}}{1-\frac{Y_{i\widetilde{\omega }-1}(2i\sqrt{z_{m}})}{Y_{i%
\widetilde{\omega }+1}(2i\sqrt{z_{m}})}}\right] .  \label{53}
\end{equation}

From Eqs. (40) and (41), we find

\begin{equation}
\frac{J_{i\widetilde{\omega }+1}(2i\sqrt{z_{m}})}{J_{i\widetilde{\omega }%
-1}(2i\sqrt{z_{m}})}\equiv \frac{Y_{i\widetilde{\omega }-1}(2i\sqrt{z_{m}})}{%
Y_{i\widetilde{\omega }+1}(2i\sqrt{z_{m}})}\sim O\left( z_{m}\right) ,
\label{54}
\end{equation}

in the NH region. Thus, the resonance condition (53) becomes

\begin{eqnarray}
\tan (i\widetilde{\omega }\pi ) &\sim &-\frac{J_{i\widetilde{\omega }-1}(2i%
\sqrt{z_{m}})}{Y_{i\widetilde{\omega }+1}(2i\sqrt{z_{m}})},  \notag \\
&=&-i\frac{\pi e^{-\pi \widetilde{\omega }}}{\widetilde{\omega }\Gamma
^{2}\left( i\widetilde{\omega }\right) }z_{m}^{i\widetilde{\omega }}.
\label{55}
\end{eqnarray}

One can immediately realize from Eq. (48) that the resonance conditions (53)
and (55) are small quantities. We can therefore use an iteration scheme to
resolve the resonance conditions. The $0^{th}$ order resonance equation is
given by \cite{Hodnew}:

\begin{equation}
\tan (i\widetilde{\omega }_{n}^{(0)}\pi )=0,  \label{56}
\end{equation}

which implies that

\begin{equation}
\widetilde{\omega }_{n}^{(0)}=-in,\text{ \ \ \ \ \ \ \ \ \ }(n=0,1,2.......).
\label{57}
\end{equation}

The $1^{st}$ order resonance condition is obtained after substituting
Eq.(57) into r.h.s of Eqs. (53) and (55). Hence, we have

\begin{equation}
\tan (i\widetilde{\omega }_{n}^{(1)}\pi )=\pm i\frac{\pi e^{i\pi n}}{\left(
-in\right) \Gamma ^{2}\left( n\right) }z_{m}^{n},  \label{58}
\end{equation}

which reduces to

\begin{equation}
\tan (i\widetilde{\omega }_{n}^{(1)}\pi )=\mp n\frac{\pi \left(
-z_{m}\right) ^{n}}{\left( n!\right) ^{2}},  \label{59}
\end{equation}

where minus (plus) stands for the DBC (NBC). For having the general
characteristic resonance spectra of the caged GHSBH, we use the fact that

\begin{equation}
\tan (x+n\pi )=\tan (x)\approx x,  \label{60}
\end{equation}

in the $x\ll $ regime. Namely, we obtain

\begin{equation}
i\widetilde{\omega }_{n}\pi =n\pi \mp n\frac{\pi \left( -z_{m}\right) ^{n}}{%
\left( n!\right) ^{2}}.  \label{61}
\end{equation}

Therefore, one finds

\begin{equation}
\widetilde{\omega }_{n}=-in\left[ 1\mp \frac{\left( -z_{m}\right) ^{n}}{%
\left( n!\right) ^{2}}\right] .  \label{62}
\end{equation}

From Eq. (38), we read the BQNMs as

\begin{equation}
\omega _{n}=-i\kappa n\left[ 1\mp \frac{\left( -z_{m}\right) ^{n}}{\left(
n!\right) ^{2}}\right] ,\text{ \ \ \ \ }(n=0,1,2.......).  \label{63}
\end{equation}

Here, $n$ is called the overtone quantum number or the so-called resonance
parameter \cite{Hod3}. For the highly excited states ($n\rightarrow \infty $
), Eq. (63) behaves as

\begin{equation}
\omega _{n}\approx -i\kappa n.\text{ \ \ \ \ \ \ \ \ \ \ \ \ (}n\rightarrow
\infty \text{).}  \label{64}
\end{equation}

The above result is in accordance with the results of \cite%
{Hodnew,Nollert,Hod08,Hod09,Hod10}. Hence, the transition frequency becomes

\begin{equation}
\Delta \omega _{I}=\kappa =\frac{2\pi T_{H}}{\hbar }.  \label{65}
\end{equation}

Substituting Eq. (65) into Eq. (3), we obtain

\begin{equation}
I_{adb}=\frac{S^{BH}}{2\pi }\hbar .  \label{66}
\end{equation}

Acting upon the Bohr-Sommerfeld quantization rule ($I_{adb}=\hbar n$), we
find the entropy spectrum as

\begin{equation}
S_{n}^{BH}=2\pi n.  \label{67}
\end{equation}

Furthermore, since $S^{BH}=\frac{\mathcal{A}}{4\hbar },$ we can also read
the area spectrum:

\begin{equation}
\mathcal{A}_{n}=8\pi \hbar n.  \label{68}
\end{equation}

Thus, the minimum area spacing becomes

\begin{equation}
\Delta \mathcal{A}_{\min }=8\pi \hbar .  \label{69}
\end{equation}

which represents that the entropy/area spectra of the GHSBH are evenly
spaced. It is obvious that the spectra of the GHSBH are clearly independent
of the dilaton parameter $a$. Furthermore, the spectral-spacing coefficient
becomes $\epsilon =8\pi $, which is in agreement with the Bekenstein's
original result \cite{Bek3,Bek4,Bek5}. In short, our result (69) supports
the study of Wei et al. \cite{Wei}.

\section{Conclusion}

In this paper, the quantum entropy/area spectra of the GHSBH are
investigated via the MM that is based on the adiabatic invariant formulation
(3) of the BHs. For this purpose, we have considered caged GHSBH whose
confining cage (mirror) is placed in the NH region $r_{m}\approx r_{+}$ \cite%
{Hodnew}. We have therefore attempted to find the BQNMs (resonance spectra)
of the GHSBH. The massless KGE for the GHSBH geometry has been separated
into the angular and the radial parts. In particular, the Zerilli equation
(21) with its effective potential (26) of the associated radial equation has
been obtained. The NH form of the Zerilli equation is well approximated by a
Bessel differential equation. After imposing the boundary conditions
appropriate for purely ingoing waves at the event horizon with the DBC and
NBC, we have obtained the resonant frequencies of the caged GHSBH. We have
then applied the MM to the highly damped BQNMs to derive the entropy/area
spectra of the GHSBH. The obtained spectra are equally spaced and are
independent of the physical parameters of the GHSBH as concluded in the
study of Wei et al. \cite{Wei}. Moreover, our results support the Kothawala
et al.'s conjecture \cite{Samp1}, which states that the BHs in Einstein's
gravity theory have equispaced area spectrum.


\begin{thebibliography}{99}
\bibitem{Robson} J. Robson, \textit{Keeping It Real (Quantum Gravity, Book 1)%
}, (Pyr Press, London, 2006).

\bibitem{Oerter} R. Oerter, \textit{The Theory of Almost Everything: The
Standard Model, the Unsung Triumph of Modern Physics}, (Plume, New York,
2006).

\bibitem{Hawking1} S.W. Hawking, Nature (London) \textbf{248}, 30 (1974).

\bibitem{Hawking2} S.W. Hawking, Commun. Math. Phys. \textbf{43}, 199 (1975).

\bibitem{Bek1} J.D. Bekenstein, Lett. Nuovo Cimento \textbf{4}, 737 (1972).

\bibitem{Bek2} J.D. Bekenstein, Phys. Rev. D \textbf{7}, 2333\ (1973).

\bibitem{Bek3} J.D. Bekenstein, Lett. Nuovo Cimento \textbf{11}, 467 (1974).

\bibitem{Bek4} J.D. Bekenstein, arXiv: gr-qc/9710076 (1997).

\bibitem{Bek5} J.D. Bekenstein, arXiv: gr-qc/9808028 (1998).

\bibitem{Gua} G. Gua, Gen. Relativ. Gravit. \textbf{45}, 1711 (2013).

\bibitem{Maggiore} M. Maggiore, Phys. Rev. Lett. \textbf{100}, 141301 (2008).

\bibitem{Kunstatter} G. Kunstatter, Phys. Rev. Lett. \textbf{90}, 161301
(2003).

\bibitem{Vagenas} E.C. Vagenas, J. High Energy Phys. \textbf{0811}, 073
(2008).

\bibitem{Medved} A.J.M. Medved, Class. Quantum Grav. \textbf{25}, 205014
(2008).

\bibitem{Sakalli14} I. Sakalli, arXiv: 1406.5130 (2014).

\bibitem{Hod1} S. Hod, Phys. Rev. Lett. \textbf{81,} 4293 (1998).

\bibitem{Hod2} S. Hod, Phys. Rev. D \textbf{59}, 024014 (1998).

\bibitem{Samp1} D. Kothawala, T. Padmanabhan and S. Sarkar, Phys. Rev. D 
\textbf{78}, 104018 (2008).

\bibitem{Samp2} W. Li, L. Xu and J. Lu, Phys. Lett. B \textbf{676}, 117
(2009).

\bibitem{Samp3} A.L. Ortega, Phys. Lett. B \textbf{682}, 85 (2009).

\bibitem{Samp4} M.R. Setare and D. Momeni, Mod. Phys. Lett. A \textbf{26},
151 (2011).

\bibitem{Samp5} C.Z. Liu, Eur. Phys. J. C \textbf{72}, 2009, (2012).

\bibitem{Samp6} S. Sebastian and V.C. Kuriakose, Mod. Phys. Lett. A \textbf{%
28}, 1350149 (2013).

\bibitem{Samp7} I. Sakalli, Mod. Phys. Lett. A \textbf{28}, 1350109 (2013).

\bibitem{Samp8} I. Sakalli and S.F. Mirekhtiary, Astrophys. Space Sci. 
\textbf{350}, 727 (2014).

\bibitem{GHS} D. Garfinkle, G.T. Horowitz and A. Strominger, Phys. Rev. D 
\textbf{43}, 3140 (1991). Erratum: \textbf{45}, 3888 (1992).

\bibitem{Chandra} S. Chandrasekhar, \textit{The Mathematical Theory of Black
Holes}, (Oxford University Press, USA, 1998).

\bibitem{Wei} S.W. Wei, Y.X. Liu, K.Yang and Y. Zhong, Phys. Rev. D \textbf{%
81}, 104042 (2010).

\bibitem{Chen} S. Chen and J. Jing, Class. Quantum Grav. \textbf{22}, 535,
(2005).

\bibitem{Motl} L. Motl and A. Neitzke, Adv. Theor. Math. Phys. \textbf{7},
307 (2003).

\bibitem{Berti} E. Berti, arXiv: gr-qc/0411025 (2004).

\bibitem{Hodnew} S. Hod, Eur. Phys. J . C \textbf{74}, 3137 (2014).

\bibitem{Cardoso} V. Cardoso, O. J. C. Dias, J. P. S. Lemos and S. Yoshida,
Phys. Rev. D \textbf{70}, 044039 (2004); Erratum: D \textbf{70},049903
(2004).

\bibitem{RanLi1} R. Li and J. Zhao, Eur. Phys. J. C \textbf{74}, 3051 (2014).

\bibitem{Herdeiro1} C.A.R. Herdeiro, J.C. Degollado and H.F. R\'{u}narsson,
Phys. Rev. D\textbf{\ 88}, 063003 (2013).

\bibitem{Herdeiro2} J.C. Degollado and C.A.R. Herdeiro, Phys. Rev. D \textbf{%
89}, 063005 (2014).

\bibitem{RanLi2} R. Li and J. Zhao, Phys. Lett. B \textbf{740,} 317 (2015).

\bibitem{RanLi3} R. Li, J. Zhao, X. Wu and Y. Zhang, arXiv: 1501.07358
(2015).

\bibitem{Okawa14} H. Okawa, V. Cardoso and P. Pani, Phys. Rev. D \textbf{90}%
, 104032 (2014).

\bibitem{Abram} M. Abramowitz and I.A. Stegun, \textit{Handbook of
Mathematical Functions}, (Dover, New York, 1965).

\bibitem{Wald} R.M. Wald, \textit{General Relativity} (The University of
Chicago Press, Chicago and London, 1984).

\bibitem{NullGeo} S. Fernando, Phys. Rev. D \textbf{85}, 024033 (2012).

\bibitem{Lambert} R.M. Corless, G.H. Gonnet, D.E.G. Hare, D.J. Jeffrey and
D.E. Knuth, Adv. Comput. Math. \textbf{5}, 329 (1996).

\bibitem{Olver} F.W.J. Olver, D.W. Lozier, R.F. Boisvert and C.W. Clark, 
\textit{NIST} \textit{Handbook of Mathematical Functions}, (Cambridge
University Press, New York, 2010).

\bibitem{Hod3} S. Hod, Phys. Rev. D \textbf{90,} 024051 (2014).

\bibitem{Nollert} H-P. Nollert, Phys. Rev. D \textbf{47,} 5253 (1993).

\bibitem{Hod08} S. Hod, Phys. Rev. D \textbf{78}, 084035 (2008).

\bibitem{Hod09} S. Hod, Phys. Rev. D \textbf{80}, 064004 (2009).

\bibitem{Hod10} S. Hod, Phys. Lett. A \textbf{374}, 2901 (2010).
\end{thebibliography}
\end{document}